\documentstyle[11pt,epsfig,amssymb]{article}
\newlength{\dinwidth}
\newlength{\dinmargin}
\newcommand {\nn} {\nonumber}
\newcommand {\half} {\frac{1}{2}}
\newcommand {\p} {\prime}
\newcommand {\G} {{\cal G}}

\newcommand{\be}{\begin{equation}}
\newcommand{\ee}{\end{equation}}
\newcommand{\bea}{\begin{eqnarray}}
\newcommand{\eea}{\end{eqnarray}}
\newcommand{\bastar}{\begin{eqnarray*}}
\newcommand{\eastar}{\end{eqnarray*}}
\newcommand{\JHEP}{\em J. High Energy Physics}

\begin{document}
\thispagestyle{empty} \addtocounter{page}{-0} \centerline{\Large
\bf Completely localized gravity with higher curvature terms}
\vspace{0.2cm}

\vspace*{0.8cm}
\centerline{ \bf Ishwaree P Neupane}
\vspace*{0.4cm}

\centerline{\sl The Abdus Salam ICTP, Strada Costiera, 11-34014, Trieste,
Italy}
\centerline{and}
\centerline{\sl School of Physics, Seoul National University,
151-747, Seoul, Korea}
%%\centerline{ E-mail: ishwaree@ictp.trieste.it }
%%\centerline{Received ....}
%%\centerline{Published ...}
%%\centerline{Online at stacks.iop.org/QCG/...}

\vspace*{0.5cm}
%%\begin{flushleft}
\centerline{\bf Abstract}
%%\end{flushleft}

\vspace*{0.5cm}

In the intersecting braneworld models, higher curvature
corrections to the Einstein action are necessary to provide a
non-trivial geometry (brane tension) at the brane junctions. By
introducing such terms in a Gauss-Bonnet form, we give an
effective description of localized gravity on the singular
delta-function branes. There exists a non-vanishing brane tension
at the four-dimensional brane intersection of two $4$-branes.
Importantly, we give explicit expressions of the graviton
propagator and show that the Randall-Sundrum single-brane model
with a Gauss-Bonnet term in the bulk correctly gives a massless
graviton on the brane as for the RS model. We explore some crucial
features of completely localized gravity in the solitonic
braneworld solutions obtained with a choice ($\xi=1$) of
solutions. The no-go theorem known for Einstein's theory may not
apply to the $\xi=1$ solution. As complementary discussions, we
provide an effective description of the power-law corrections to
Newtonian gravity on the branes or at the common intersection
thereof.

\vspace*{0.5cm}

{PACS} numbers: 04.50.+h, 11.10.Kk, 11.25.Mj
%\end{flushleft}

\vspace*{0.6cm}

\baselineskip=18pt

%%\newpage

\setcounter{equation}{0}

%=========================================================
\section{Introduction}
\label{sec-intro}

In recent years, it has been understood that the brane-world
models of non-compact extra dimensions~\cite{RS2} reproduce the
usual Einstein gravity. This idea was quickly generalized to a
discrete patch of $AdS_{4+N}$ space accommodating $N$ intersecting
$(N+2)$ branes~\cite{Hamed}. Later, the RS singular $3$-brane
model with perturbations of bulk geometry was fully investigated
in~\cite{GTT,GKR,Csaki00a}, and interesting discussion of the
'crystal' multi-braneworld solutions appeared in~\cite{Kalo}. A
class of {\it solitonic} brane solutions arising from a special
fine tuning between the bulk cosmological term and the
Gauss-Bonnet coupling was given in~\cite{CZK}. The behavior of
gravity localized on branes or string-like defects in $N\geq 1$
dimensions was studied in~\cite{Gherghetta}. These constructions
have been useful in providing some realizations of trapped
four-dimensional gravity to a brane embedded in an AdS space. It
could be possible to localize ordinary matter and gauge fields to
this $(3+1)$ dimensional manifold if the RS brane is a
$D3$-brane-like object of string/M theory. Since a $D3$-brane in
the string theory background is necessarily a solitonic object,
and only the higher curvature corrections can provide a
non-trivial background at the brane junction, we find it
interesting to study intersecting brane-world configurations by
including a Gauss-Bonnet term.

In the intersecting brane configurations
of~\cite{Hamed,Kalo,CCJ,Csaki00a}, a $3$-brane given by the
four-dimensional intersection of $(2+N)$-branes ($N\geq2$) with
larger co-dimensions has been viewed as our universe. In this
picture, though each extra dimension has an infinite extension,
the effective size of $N$ non-compact extra dimensions is supposed
to be finite, which could be as large as $ 1 \mu m$ (for $N\geq
2$) or much larger than their natural scale $\ell
>>M_{Pl}^{-1}~(\sim 10^{-33} cm)$~\cite{Hamed1,Hamed}. While
taking a Gauss-Bonnet coupling into account, the effective
four-dimensional Planck mass reads $M_{eff}^2=M_*^{2+N}\,V_N$,
where the effective volume of $N$ extra dimensions is \be
V_N=\left(\frac{2}{k}\right)^N\, \frac{(1+\xi)}{(N+1)!}\,. \ee
Here $\xi$ is a coupling constant arising from a GB term, and the
constant $k$ has dimension of inverse AdS length. One of the
motivations here to include a GB coupling is that the low energy
version of certain string theories (such as type I heterotic
strings) admits the quadratic curvature corrections as a GB
invariant in their effective actions~\cite{Metsaev87a}. It is
remarkable that the inclusion of a GB term can consistently
explain the basic features of the Randall-Sundrum type brane-world
models in $D=5$~\cite{IPN01a,HMLee,CNW}. Some other aspects of
Randall-Sundrum brane models supplemented with higher curvature
terms were studied in recent papers,
including~\cite{Zee,Rizos,KKL,KAKU,Olecho1,Olecho2,Giovannini01a}.
Here we construct brane-world solutions for $N\geq 1$ with a
Gauss-Bonnet term. Indeed, the inclusion of a GB term in the bulk
might shed more light towards the understanding of the RS models
in the $AdS_{4+N}$ spaces. One of the interesting observations is
that a GB term does allow a non-trivial four-dimensional brane
tension when there are two extra transverse directions. Also a
Gauss-Bonnet term could be useful to circumvent some of the
arguments given for the {\it no-go
theorem}~\cite{Freedman99a,Malda00a} applicable to the braneworld
models based on the Einstein action.

The rest of paper is organized as follows. In section 2, we give a
general set-up of the effective braneworld action, and present a
framework for intersecting braneworld configurations with a GB
term. Section 3 explains the basic features of the linear
perturbation equations in $(4+N)$ dimensions. In section 4, we
derive a relation between the effective four-dimensional Planck
mass and $(4+N)$ dimensional fundamental mass scale. In section 5,
we give the expressions for graviton propagators in $AdS_{4+N}$
space, and discuss some salient features of gravity localized on
the branes. In section 6, we elucidate upon the braneworld
solutions in five dimensions, which exhibit that $\xi=1$ is indeed
a physical choice. We then give in section 7 a complementary
description of Newton potential corrections for solitonic branes.
In section 8, we briefly comment upon the scalar and vector modes
of the metric fluctuations. Section 9 contains conclusion.

%%%%%%%%%%%%%%%%%%%%%%%%%%%%%%%%%%%%%%%%%%%%%%%%%%%%%%%%%%%%%%%%%%%%%
%%%%%%%%%%%%%%%%%%%%%%%%%%%%%%%%%%%%%%%%%%%%%%%%%%%%%%%%%%%%%%%%%
\section{Intersecting branes setup}
In the low energy effective action of the Horova-Witten type I
heterotic string theory~\cite{Horava1}, there arise additional
interactions of the gauge fields and higher powers of the
curvatures. In particular, there is $R^2$ interaction in the
Gauss-Bonnet combination~\cite{Kashima}. In many cases a
contribution of higher curvature terms is neglected as it is
higher order in derivatives, but in the warped braneworld scenario
in $D\geq 5$, we see the relevance of these terms if they are
expressed in a Gauss-Bonnet form. We will not stick here to the
physical origin of a Gauss-Bonnet term. Rather, we will try to
explore the physical importance of this term in warped
backgrounds. We shall start with the following braneworld
effective action in $D(=(3+N)+1)$-dimensional space-time $({\cal
B} )$, where $\partial {\cal B}$ represents the
$(3+N)$-dimensional boundary,
\begin{eqnarray}\label{action}
S&=&\int_{{\cal B}} d^{4+N}x\,\sqrt{-g_{4+N}}\,
\left(\frac{R}{\kappa}-2\Lambda
+\alpha\left(R^2-4R_{PQ}R^{PQ}+R_{MNPQ}R^{MNPQ}\right)\right)\nn\\
&{}&+\sum_{i=1}^N\int_{\partial {\cal B}}^{i'th~brane}
d^{3+N}x\sqrt{-g_{3+N}}\,\left(-\Lambda_i\right)\nn \\
&{}&+\sum_{i\neq j}^N\int^{i,j'th~brane}
d^{2+N}x\sqrt{-g_{2+N}}\,\left(-\Lambda_{i,j}\right)\nn \\
&{}&+\int d^4 x \sqrt{-g^{z_1=0,\cdots,z_n=0}}\,
(-\lambda)\,.
\end{eqnarray}
Here $P,Q, \cdots=0,1,\cdots, 3+N$, $x^{\mu}(\mu = 0,...,3)$
parametrize brane coordinates with Lorentzian signatures and $z^i
(i= 1,...,N)$ parametrize extra non-compact dimensions. The
action~(\ref{action}) characterizes an array of $N$ orthogonal
$(N+2)$-spatial dimensional branes in $(4+N)$ dimensions.
$\Lambda$ is the $D$-dimensional bulk cosmological term,
$\Lambda_{i}$ are the brane tensions of the $N$ intersecting
$(N+2)$-branes, $\Lambda_{i,j}$ are $(N+1)$-brane tensions, and
$\lambda$ is the four-dimensional brane tension at the common
intersection of higher dimensional branes. The coupling $\alpha$
has the mass dimension of $M_*^{N}$, and $\kappa=16\pi
G_{4+N}\equiv M_*^{2+N}$.

The graviton equations of motion derived by varying the
action~(\ref{action}) with respect to $g^{MN}$ take the following
form
\begin{eqnarray}
&&\sqrt{-g_{4+N}}\left(G_{MN}+
\kappa\,H_{MN}+\kappa\,\Lambda g_{MN}\right)=
-\frac{\kappa}{2}\,\sum_{i=1}^N \Lambda_i\,
\sqrt{-g_{3+N}^{(z_i=0)}}\,\delta(z_i)\,
\delta_M^p\delta_N^q g_{pq}^{(z_i=0)}\nn \\
&&~~~~~~~~~~~~~-\frac{\kappa}{2}\sum_{i\neq j}^N\Lambda_{i,j}\,
\sqrt{-g_{2+N}^{(z_i,z_j=0)}}\,\delta(z_i)\delta(z_j)\,
\delta_M^r\delta_N^s g_{rs}^{(z_i,z_j=0)} \nn \\
&&~~~~~~~~~~~~-\frac{\kappa\,\lambda}{2}\,
\sqrt{- g_{4}\,^{(z_1,z_2,\cdots,z_N=0)}}
\,\delta(z_1)\delta(z_2)\cdots
\delta(z_N)\delta_M^\mu\,\delta_N^\nu\,g_{\mu\nu}^{z_1,z_2,\cdots,z_N=0}\,,
\label{eqofmotion}
\end{eqnarray}
where $H_{MN}$, an analogue of the Einstein tensor stemmed from
the GB term, reads
\begin{eqnarray}
H_{MN}&=&-\frac{\alpha}{2}\, g_{MN}
\left(R^2-4R_{PQ}R^{PQ}+ R_{PQRS}R^{PQRS}\right)\nn\\
&{}&+2\alpha \left[R R_{MN}-2R_{MPNQ}R^{PQ}+
R_{MPQR}R_N\,^{PQR}-2R_M\,^P R_{NP}\right]\,.
\end{eqnarray}
In equation~(\ref{eqofmotion}), the indices $(p,q)$ take $(3+N)$
possible values, while $(r,s)$ take only $(2+N)$ possible values,
and $\mu,\nu=0,1,2,3$. The $(2+N)$-brane tensions $\Lambda_i$
multiply only $\delta(z_i)$ (i.e. the terms involving only one
delta function), $\Lambda_{i,j}$ multiply two delta functions, and
the brane tension at the common intersection $(\lambda)$
multiplies a product of delta functions
$\delta(z_1),\cdots,\delta(z_N)$. In $D=7$, for example, $\lambda$
multiplies $\delta(z_1)\delta(z_2)\delta(z_3)$. One observes that
when $N=2$, $\Lambda_{1,2}$ is replaced by $\lambda$, so that
$\lambda$ can have a non-zero value at the intersection of two
$4$-branes as we exhibit below.

We write a $(3+N)+1$-dimensional metric as
%%%%%%%%%%%%%%%%%%%%%%%%%%%%%%%%%%%%%%%%%%%%%%%%%%%%%%%%%%
\begin{equation}
ds^2= e^{-2A(z)} \left(g_{\mu\nu}(x^\lambda)dx^{\mu}dx^{\nu}
+g_{ij}(z)dz^i dz^j\right)\,.
\label{genmetric}
\end{equation}
We assume that the four-dimensional spacetime is Minkowski. Then
we may express the metric~(\ref{genmetric}) in the following form,
which represents the Poincar\'e half parametrization of
$AdS_{4+N}$ space~\cite{Hamed},
\begin{equation}
ds_{4+N}^2=
\frac{\ell^2}{\bar{z}^2}\,\big(\eta_{\mu\nu}dx^{\mu}dx^{\nu} +
d\bar{z}^2+ d\Omega_{N-1}^2\big)\,. \label{conmetric}
\end{equation}
The length scale $\ell$ is fixed by the $(4+N)$-dimensional bulk
cosmological term $\Lambda$. From the exact non-linear analysis in
$(4+N)$-dimensions we can find the most general solution to
modified Einstein field equations, admitting a normalisable $4d$
graviton. This solution is $A(z)=\log\big(k_i|z_i|+1\big)$, where
$z_i=z_1,\,z_2,\cdots$ count extra spaces and
$k_i=k_1,\,k_2,\cdots$ are the inverse of $AdS$ curvature radii.
The choice $k_1=k_2=\cdots=k$ further keeps the
metric~(\ref{conmetric}) manifestly symmetric under permutations
of all extra $N$ dimensions, and the length scale $L$ ($\equiv
(V_N)^{1/N}$) can be interpreted as the compactification size of
$N$ extra dimensions~\cite{Hamed}. The metric solution then
becomes
\begin{equation}\label{mainsol}
ds_{4+N}^2= \frac{1}{(k\sum_{i=1}^N\,|z_i|+1)^2}\,
\Big(\eta_{\mu\nu}dx^{\mu}dx^{\nu} +\sum_{j=1}^N\,(d
z^j)^2\Big)\,. \label{conmetric2}
\end{equation}
Here $k\equiv (\sqrt{n}\,\ell)^{-1}$. Since the coordinates $z^j$
parametrize the extra dimensions, the full bulk space ($z^j\neq
0$) comprises $2^N$ identical patches of the $AdS_{4+N}$ space,
which are glued together along the branes ($z^j=0$). Here we have
taken AdS space as the simplest choice of the bulk spacetime. And
branes are characterized by the delta-function-like singularities,
which arise from the second derivative of the warp factor $A(z)$.
However, there might exist other possibilities of an AdS bulk,
such as the AdS Schwarzschild black-hole spacetime considered
in~\cite{Nojiri007}. It is also plausible that within the AdS/CFT
correspondence gravity can be trapped on the brane in a manner
similar to that in the original RS scenario~\cite{RS2}, when the
brane is embedded into the AdS Schwarzschild
bulk~\cite{Cvetic01a}. It would be interesting to study the
gravitational perturbations in such a black-hole background.

We have assumed that the warp factor $A(z)$ is a function of all
transverse coordinates, that is, $z^i$ \be
A(z)=\log\left(\sum_{i=1}^{N} k\,|z_i|+1\right) \,. \ee Then a
straightforward calculation yields \bea
{A'}^2&=&e^{-2A(z)}\,k^2\,\sum_{i}^N
\left(\partial_{z_i}|z|\right)^2 =e^{-2A(z)}\,k^2\,N
\,,\label{oneprime}\\
A^{\p\p}&=& -e^{-2A(z)}\,k^2\,N+e^{-A(z)}\,2 k
\sum_{i}^N\delta(z_i)\,, \label{twoprimes} \eea where primes
represent differentiation with respect to $z$. The metric
solution, in terms of the warp factor $A(z)$, therefore satisfies
a relation $A^{\p\p}+{A^\p}^2=e^{-A(z)}\,2k\,\delta(z)\geq 0$.
Here, we simply note that this condition is equivalent to the weak
energy condition one derives from
$G_t^t-G_z^z=\kappa\left(T_t^t-T_z^z\right)=(N+2)\left(A^{\p\p}+{A^\p}^2\right)\geq
0$.

In order to exhibit localized behavior of gravity to the brane
intersections, one may look at the linear perturbations about the
background given by~(\ref{conmetric2}). For this purpose, we
perform a conformal transformation $g_{MN}=e^{-2A(z)} {\tilde
g}_{MN}$, where $A(z)=\log(k\sum_i|z_i|+1)$. Then, we adopt a
background subtraction technique as in~\cite{CCJ}. Being specific,
we use the linear equation
$\delta\hat{G}_{MN}+\kappa\,\delta\hat{H}_{MN}=0$, where
$\delta\hat{G}_{MN}=\delta G_{MN}-\bar{G}_{MP} h_N\,^P$,\,
$\delta\hat{H}_{MN}=\delta H_{MN}-\bar{H}_{MP}\,h_N\,^P$. Note
that $\bar{G}_{MN}=\kappa\,\bar{T}_{MN}$ gives the background
Einstein equations. What we are equating here with $\bar{G}_{MN}$
are only the vacuum energy contributions to the energy-momentum
tensor. In our approach, the Randall-Sundrum-type fine tunings
will only be implicit. If we wish, we can see them by considering
the difference between $(\delta G_{\mu\nu}+\kappa\,\delta
H_{\mu\nu})$ and $(\delta
\hat{G}_{\mu\nu}+\kappa\,\delta\hat{H}_{\mu\nu})$. This gives
after some simplifications,
\begin{eqnarray}
&&-\frac{1}{2}\Big[\bar{R}\,h_{\mu\nu}-2\bar{R}_{\mu}\,^\lambda\,
h_{\nu \lambda}\Big]
-\frac{\alpha\kappa}{2}\Big[\left(\bar{R}^2-4\bar{R}_{MN}^2
+\bar{R}_{MNPQ}^2\right)h_{\mu\nu}\nn \\
&&~~-4\left(\bar{R}\bar{R}_\mu\,^\lambda-2\bar{R}_\mu\,^{P\lambda Q}
\bar{R}_{PQ}-2\bar{R}_{\mu S}\bar{R}^{\lambda S}
+\bar{R}_{\mu PQR}\,\bar{R}^{\lambda PQR}\right)
\,h_{\nu\lambda}\Big]\,.\label{extraterms}
\end{eqnarray}
One could fine-tune the contributions coming from these terms to
the vacuum energy contributions on the branes, that is, the source
terms on the right-hand side of equation~(\ref{eqofmotion}). For
this purpose, we parameterize $h_{\mu\nu}=e^{(N+2)
A(z)/2}\,\tilde{h}_{\mu\nu}$ and obtain
\begin{eqnarray}
&&\Big[N(N+2)(N+3)\,k^2\,e^{-2A}\,(1-\xi/2)-
8(N+2) k\, e^{-A}(1-\xi/3)\,\sum_{i=1}^{N}\delta(z_i)\nn \\
&&-16(N+1)(N+2)\,\alpha\,\kappa\,
k^2 \sum_{i\neq j}\delta(z_i)\delta(z_j)\,\Big]\tilde{h}_{\mu\nu}(x,z)=\nn\\
&&-2\kappa \left[e^{-2A(z)}\Lambda+e^{-A(z)}\sum_{i=1}^{N}\Lambda_{i}\,
\delta(z_i)+\sum_{i\neq j}\Lambda_{i,j}\,
\delta(z_i)\delta(z_j)\right]\,\tilde{h}_{\mu\nu}(x, z)\,,
\end{eqnarray}
where $\xi=2N^2(N+1)\alpha'\kappa\,M_*^{-2}k^2$ and we have
replaced $\alpha$ by $M_*^{N}\alpha'$, so that $\alpha'$ is a
dimensionless Gauss-Bonnet coupling. Here, it might be relevant to
note that in the full non-linear analysis the fine-tuned relations
emerge as boundary conditions to be satisfied on the brane(s),
while the bulk part amounts to express $\Lambda$ in terms of
$k^2$; see~\cite{KKL} for the $N=2$ case. For arbitrary $N$, we
obtain
\begin{eqnarray}
\Lambda&=&-\frac{N(N+2)(N+3)k^2\,(1-\xi/2)}
{2\,\kappa}\label{finetune1}\,,\\
\Lambda_{i}&=&\, \frac{4(N+2)\,k\,(1-\xi/3)}
{\kappa}\label{finetune2}\,,\\
\Lambda_{i,j}&=&8(N+1)(N+2)\alpha'\,M_*^2\,k^2\,.
\label{finetune3}
\end{eqnarray}
For $N=2$, one may replace the brane tension $\Lambda_{1,2}$ by
$\lambda$, because there is only one four-dimensional brane
intersection for two $4$-branes that mutually intersect. A clear
message is that the brane tension $\Lambda_{1,2}$ is non-vanishing
for $\alpha^\p>0$, but it vanishes for the Einstein gravity
($\alpha=0$).

We may perturb the background metric as \be\label{background1}
ds_{4+N}^2= \frac{1}{(k\sum_{i=1}^N\,|z_i|+1)^2}\,
\left(\left(\eta_{\mu\nu}+h_{\mu\nu}\right) dx^{\mu}dx^{\nu}
+\sum_{j=1}^N\,(d z_j)^2\right)\,. \ee In this gauge, however,
there may be some additional gravitational degrees of freedom
coming from off-diagonal components $(h_{\mu m})$ and diagonal
components $(h_{mm})$ of the perturbed equations for $h_{MN}$,
where $m$ could run from $z_1$ to $z_N$. We will comment upon
these modes in the section $8$. One may also use the approximation
where the tensor modes either decouple from the perturbed
equations for $h_{\mu m}$ and $h_{mm}$, or only the non-vanishing
components of the fluctuations are
$h_{\mu\nu}$~\cite{Hamed,Csaki00a}. In fact, in the ordinary
two-derivative gravity supplemented by a GB term there are no
extra degrees of freedom than that of Einstein's theory.

We find reasonable to study first the linear tensor fluctuations,
$h_{\mu\nu}$ in the gauge $h_\mu\,^\mu=0,\,\partial^\lambda
h_{\lambda\mu}=0$. The linearized equations for $h_{\mu\nu}$ then
take the following form \bea\label{lineareq1}
&&\frac{1}{\kappa}\,\Big[-\Box_4-\Box_z+(N+2)\,k
\,e^{-A(z)}\sum_{i=1}^N sgn(z_i)
\partial_{z_i}\Big]h_{\mu\nu}\nn \\
&&+2\alpha\, N\bigg[N(N+1)k^2\left(\Box_4+\Box_z\right)-4k\, e^{A(z)}
\sum_{i=1}^n
\delta(z_i)\Box_4 -4k\, e^{A(z)}\sum_{i\neq j}^N\delta(z_i)\partial_{z_j}^2
\nn \\
&&+(N+1)\,k^2\Big(4\sum_{i=1}^N\delta(z_i)-N(N+2)k\,
e^{-A(z)}\Big)\sum_{j=1}^N sgn(z_j)\partial_{z_j}\bigg] h_{\mu\nu}=0
\eea
where $\Box_z=\partial_{z_1}^2+\partial_{z_2}^2+\cdots \partial_{z_N}^2$.
When $\alpha=0$, our results will reproduce the results
appeared in~\cite{Hamed}. But here we find more interesting new
features which are unavailable if the gravity action does not contain
higher curvature corrections.

%%%%%%%%%%%%%%%%%%%%%%%%%%%%%%%%%%%%%%%%%%%%%%%%%%%%%%%%%%%%%%%%%%
\section{Localized gravity in $(4+N)$-dimensions}

One can remove from equation~(\ref{lineareq1}) the single (linear
in) derivative term from first square bracket by re-scaling the
metric: $h_{\mu\nu}=e^{(N+2)A(z_i)/2}\,\tilde{h}_{\mu\nu}$. We
should note, however, that single derivatives of
$\tilde{h}_{\mu\nu}$, and terms like $\delta(z)\,sgn(z)\partial_z$
still survive from the second square bracket in~(\ref{lineareq1}).
Fortunately, suitable combinations of these terms satisfy the
necessary jump condition(s) on the brane(s). In fact, such terms
do not appear if one has started only with the Einstein action,
i.e. $\alpha=0$. We can make a change of variables
$\tilde{h}_{\mu\nu}(x,z)\equiv \hat{h}_{\mu\nu}(x)\psi(z)
=\epsilon_{\mu\nu} e^{ip\cdot x}\psi(z)$, where
$\epsilon_{\mu\nu}$ is the constant polarization tensor of the
graviton wave function, and $p^2=-q^2\equiv m^2$. Then, to the
leading order, \be S\sim \int d^N z\, |\psi(z)|\,^2\, \int d^4x \,
\partial^\lambda\hat{h}_{\mu\nu}(x)
\,\partial_\lambda \hat{h}^{\mu\nu}(x)+\cdots\,.
\ee
%%$$S\sim M_*^{n+2}\int d^n z\,e^{-(n+2)A(z)}\int d^4x\,%%%%
%%\sqrt{-\hat{g^{(4)}}}\,\left(R^{(4)}+\cdots\right)\,.$$ %%%%%
This actually implies that (i) $\psi(z)$ is the conventional
quantum-mechanical wavefunction, and (ii) if we had not chosen the
coordinates in a conformally flat form, the kinetic terms in the
$x$ and $z$ directions would have had different conformal factors,
but the redefinition of the metric
$h_{\mu\nu}=e^{(N+2)A(z_i)/2}\,\tilde{h}_{\mu\nu}$ puts the action
in the canonical form by absorbing the conformal factors. This is
similar to what happens in a model without the Gauss-Bonnet
interaction term~\cite{Csaki00a}.

The metric fluctuations in terms of $\tilde{h}_{\mu\nu}$ take the
following form \bea\label{lineareq2}
&&\frac{1}{\kappa}\,\left[-\Box_4-\Box_z-(N+2)k\, e^{-A(z)}\sum
\delta(z_i)
+\frac{N(N+2)(N+4)k^2}{4}\, e^{-2A(z)}\right]\tilde{h}_{\mu\nu}\nn \\
&&+N\,\alpha e^{2A(z)}\Bigg[2N(N-1)k^2 e^{-2A(z)}
\left(\Box_4+\Box_z\right)
-8k e^{-A(z)}\sum_i\delta(z_i)\,\Box_4 \nn \\
&& +8\,k^2e^{-2A(z)}\Big((N+1)\sum_i sgn(z_i)\delta(z_i)\partial_{z_i}
-\sum_{i\neq j}\delta(z_j)sgn(z_i)\partial_{z_i}\Big)\nn \\
&&-8k e^{-A(z)}\sum_{i\neq j}\delta(z_i)\partial_{z_j}^2
-4(N+2)\,k^2 e^{-2A(z)}
\Bigg(\frac{N^2(N+1)(N+4)}{8}\, k^2e^{-2A(z)}\nn \\
&&+2\sum_{i\neq j}\delta(z_i)\,\delta(z_j)
-\left(N^2+2\right)\,e^{A(z)}\sum_i\delta(z_i)\Bigg)
\Bigg]\tilde{h}_{\mu\nu}=0\,. \eea Here we have defined $\Box_4
\tilde{h}_{\mu\nu}=m^2\,\tilde{h}_{\mu\nu}$. In terms of
$\psi(z)$, we may express the above equation in the form
\begin{eqnarray}\label{linearxi}
&&(1-\xi)\,
\bigg[-m^2-\Box_z-(N+2)\,k\, e^{-A(z)}\sum_{i=1}^{N}\delta(z_i)
+\frac{N(N+2)(N+4)\,k^2}{4\,(k\sum_{i}|z_i|+1)^2}\bigg]\,\psi(z)\nn \\
&&+8 N\,\alpha\,\kappa\,k\, e^{A(z)}
\bigg[-\sum_{i=1}^{n}\delta(z_i)\,m^2- \sum_{i\neq j}^{n}\delta(z_i)\,
\partial_{z_j}^2- e^{A(z)}\,k
\Big(\sum_{i\neq j}\delta(z_j) sgn(z_i)\partial_{z_i}\nn \\
&&-(N+1)\sum_{i=1}^{N}\delta(z_i) sgn(z_i)
\partial{z_i}\Big)
+e^{2A(z)}\,\frac{(N+2)(N^2-N+4)\,k^2}{4}\sum_{i=1}^n\delta(z_i)\bigg]
\,\psi(z)\nn \\
&&~~~~~~~~~~
-8N(N+2)\,\alpha\,\kappa\,k^2\,\sum_{i\neq j}^{N}\delta(z_i)\delta(z_j)\,
\psi(z)=0\,,
\end{eqnarray}
where $\xi=2N^2(N+1)\alpha\,\kappa\,k^2$. The modified Einstein
equations in the bulk amount to giving a solution for $k^2$: \be
\label{bulk-solution} 1-\xi=\mp
\sqrt{1+\frac{8N(N+1)\,\alpha\,\kappa^2\,\Lambda} {(N+2)(N+3)}}\,.
\ee An interesting special case arises for $k=M_*/\sqrt{2N^2(N+1)
\alpha\kappa}$ (and hence $\xi=1$), where the two branches of the
bulk solution~(\ref{bulk-solution}) coincide. Of course, one
require $\alpha>0$ to ensure $k>0$. Moreover, a fine tuning
condition between the bulk cosmological term and the GB coupling
$\alpha$, \be
\Lambda=-\frac{(N+2)(N+3)}{8N(N+1)\,\alpha\,\kappa^2}\,
\label{specialtuning} \ee is useful to study the localization of
gravity to a solitonic $3$-brane~\cite{CZK}. For $N=2$, the choice
$\xi=1$ allows one to study a solitonic $3$-brane solution with
the supersymmetric interpretation~\cite{CZK}. The readers will
note that $(1-\xi)=0$ implies trivial corrections to the graviton
propagator in $(4+N)$ dimensions, since all sub-leading-order
corrections involve this factor. But one still recovers the
RS-type braneworld solutions in one lower dimensions (i.e., in
$3+N$ dimensions). In fact, the choice $\xi=1$ helps us to study
gravity confined to a solitonic brane. It is interesting that the
solutions with $\xi=1$ have structures which are similar to the
original Randall-Sundrum braneworld solutions.

\section{Finiteness of Newton's constant}
A finite four-dimensional gravitational coupling (or the Planck
mass) ensures that the ordinary four-dimensional graviton is
present as a massless and normalizable zero-mode solution of the
theory. In the RS three-brane model of co-dimension one, the four
dimensional Newton constant is determined to be \be
\frac{1}{G_N^{(4)}}\sim M_{Pl}^2 \simeq M_*^{3}\int_0^{z_c}
dz\,e^{-3A(z)}\left[1+\xi -8\alpha \kappa_5\,
e^{2A(z)}\left({A^\prime}^2-A''\right)\right] \ee where
$\xi=4\alpha\kappa_5\,e^{2A}{A'}^2$. The last term
${\left[e^{-A(z)}{A^\prime}\right]}_0^\infty$ has to be finite for
$G_N^{(4)}$ to be finite. This is finite for a solution of the
type $A(z)=\log(|z|/L+1)$, since $A^\prime(0_+)=1/L$ and the
background is invariant under $z\to -z$ symmetry.

Next consider the general case where $N$ is arbitrary. In this
case, one can find the four-dimensional Planck mass by reading off
the coefficients of $-m^2$ in the linearized expression,
equation~(\ref{lineareq2}) or equation~(\ref{linearxi}), and then
integrating over the extra dimensions. The result is  \bea
M_{Pl}^2&=&M_*^{2+N}\int_{-\infty}^{+\infty} d^N
z\,e^{-(N+2)A(z)}\Big[(1-\xi)
+8N\alpha\kappa\,k\,e^{A(z)}\sum_{i=1}^{N}\delta(z_i)\Big]\nn \\
&=& M_*^{2+N}\,\frac{2^N\,(1+\xi)}{(N+1)!}
\,k^{-N}\,.\label{planckmass} \eea We will shortly justify this
result by integrating the $4d$ part of the metric, as in the
original RS model. For this purpose, one may replace
$\eta_{\mu\nu}$ in equation~(\ref{conmetric2}) by
$\hat{g}^{(4)}_{\mu\nu}(x)$, insert the metric into the
action~(\ref{action}), and finally integrate over the $z_i$
coordinates. The net result is that \be S_{eff}={\hat
M}_{Pl}^2\int\,d^4x\,\sqrt{-\hat{g}^{(4)}}\,
\Big[\hat{R}+\hat{\alpha}\big(\hat{R}^2-4\hat{R}_{\mu\nu}^2+
\hat{R}_{\mu\nu\lambda\rho}^2\big)\Big]+\mbox{fine-tuned terms}\,,
\ee where the effective $4d$ Planck mass $\hat{M}_{Pl}$ and ${\hat
\alpha}$, the $4d$ analog of the $D$-dimensional GB coupling
$\alpha$, are defined by \bea {\hat
M}_{Pl}^2&=&M_*^{2+N}\int_{-\infty}^{+\infty} d^N
z\,e^{-(N+2)A(z)}
\Big[1-2N(N+1)(N+2)\alpha\kappa\,k^2\nn \\
&{}& +8(N+1)\alpha\,\kappa\,k\,e^{A(z)}\,\sum_{i}^{N}
\delta(z_i)\Big] \nn \\
&=&
M_*^{2+N}\,\frac{2^N}{(N+1)!}\,(1+\xi)\,k^{-N}\,,\label{planckmass2}
\eea and, \be {\hat \alpha}= \alpha\,\frac{M_*^{2+N}}{{\hat
M}_{Pl}^2}\, \int_{-\infty}^{+\infty}d^N z\,e^{-NA(z)}\,. \ee
Observe that to ensure a positivity of the four-dimensional
graviton coupling (i.e. $G_4>0$) and to preserve unitarity (i.e. a
definite positive-norm state) condition a with GB term, one must
take $(1+\xi)>0$ and $\alpha>0$.

The four-dimensional massless graviton corresponding to a bound
state with the wavefunction, up to a normalization factor, is
\begin{equation}
\psi_{bound}\sim e^{-(N+2)\,A(z)/2}\,.\label{bound}
\end{equation}
For the massless graviton mode $m^2=0$, the second square bracket
in equation~(\ref{linearxi}) then gives a non-trivial term
\begin{equation}
8N(N+2)\,\alpha\,\kappa\,k^2\,\sum_{i\neq
j}\delta(z_i)\delta(z_j)\, \psi(z)\,.
\end{equation}
This exactly cancels with the last term in
equation~(\ref{linearxi}). Thus, the massless mode of the graviton
fluctuation is unaffected by a Gauss-Bonnet action. And for $|z|>>
1/k $ one always recovers the $N$-dimensional Randall-Sundrum-type
volcano potential. This, in turn, implies that the gravity in the
bulk is not delocalized by a Gauss-Bonnet term even if $N > 1$.

%%%%%%%%%%%%%%%%%%%%%%%%%%%%%%%%%%%%%%%%%%%%%%%%%%%%%%%%%%%%
\section{Graviton propagator in $(4+N)$ dimensions}

In order to study the graviton propagator, let us set $N=1$
in~(\ref{lineareq2}) and momentarily change the bulk coordinate to
$z=e^{k|y|}/k$. Then, in terms of $A(y)(=k|y|)$, the Fourier modes
$e^{ip\cdot x}\psi(x,y)$ of the tensor fluctuation $h_{\mu\nu}$ in
five dimensions satisfy \bea
&&\Big[\big(1-4\alpha\kappa_5\dot{A}^2\big)\,
\Big(e^{2k|y|}\,m^2+\partial_{y}^2-4\dot{A}^2
+2\ddot{A}\Big)\nn \\
&& +4\alpha\kappa_5\ddot{A}\,e^{2k|y|}\,m^2
-8\alpha\kappa_5\dot{A}\ddot{A}
\big(\partial_y+2\dot{A}\big)\Big]\,\psi(y)=0\,.
\label{linear-in-y} \eea Here dot represents differentiation with
respect to $y$. To study graviton propagator for the
matter-localized brane, one may place a test mass $2\pi\,m_*$ on
the brane $z=1/k$ (or on the intersection of $(N+2)$-branes) and
ask for the corresponding Newtonian potential at a distant point
on the brane. This may be done by inserting a source term
$G_{4+N}\,(2\pi m_*)\delta^3(x)\delta(z)$ on the rhs of
equation~(\ref{linear-in-y}). The five-dimensional graviton
propagator ${\cal G}_5(x,z;x', z')$ is obtained by integrating
over the Fourier modes \bea {\cal G}_5(x,z;x', z')=\int\frac{d^4
p}{(2\pi)^4}\,
e^{ip\cdot(x-x^\prime)}\,\psi(z,z')\,.\label{fourier} \eea We also
need appropriate Neumann-type boundary conditions on the brane(s)
to proceed further. For this purpose, we need to regularize the
$\delta$-function\footnote{This issue has also been explained in
Ref.~\cite{Olecho2}, where a similar analysis is carried out by
considering arbitrary order of curvature correction terms.}. Also
note that $\partial_y\psi(0_+)=-\partial_y\psi(0_-)$, while
$\partial_y\psi(|0|)=0$. Then the last term of
equation~(\ref{linear-in-y})) gives
$$
-\frac{16\alpha \kappa_5 k^2}{3}\, \delta(y)
\left(\partial_y+2k\right)\psi(0_+)\,.
$$
Similarly, two other non-trivial terms arising from the first two
brackets in equation~(\ref{linear-in-y}), namely $-4\alpha\kappa_5
{\dot A}^2\,
\partial_y^2\psi(y)$ and $-8\alpha \kappa_5 {\dot A}^2 {\ddot A}\psi(y)$,
give
$$
-\frac{8}{3}\alpha\kappa_5 k^2\delta(y)\left(\partial_y+2k\right)\psi(0_+)\,.
\label{firstterm}
$$
The total non-trivial contribution at $y=0$, from
equation~(\ref{linear-in-y}), is therefore \be
2\left(\partial_y+2k\right)\psi(0_+) -8\alpha\kappa_5 k^2
\left(\partial_y+2k\right)\psi(0_+) +8\alpha\kappa_5 \,k\,
m^2\,\psi(0_+)\,. \ee This gives the correct Neumann-type boundary
condition on the brane after integrating from just below to the
just above the brane (i.e., in the neighborhood of $y=0$). Hence a
Neumann type boundary condition that one must impose on the brane
at $z=1/k$ is \be \Big(z\partial_z+2+\frac{\xi}{1-\xi}\,q^2
z^2\Big)\, \psi(z,z')|_{z=1/k}=0\,,\label{boundary} \ee where
$\xi=4\alpha'\,M_*^{-2} k^2$, and $m^2=q^2=-p^2$. By the same
token, one can derive $(N+4)$ dimensional Neumann boundary
condition at $z=1/k$. As the calculation is much involved, we
avoid here the technical details and simply give the final
expression \be \label{boundaryn}
\left(z\partial_{z}+\frac{N+3}{2}+\frac{2}{(N+1)}\,\frac{\xi\, q^2
z^2} {(1-\xi)}\right)\,\psi|_{z=1/k}=0\,, \ee This boundary
condition was first introduced in Ref.~\cite{GKR}, but for $\xi=0$
case. One should also satisfy the following matching conditions at
$z=z'$:
\begin{eqnarray}
\psi_{<}|_{z=z^\p}&=&\hat{\psi}_>|_{z=z^\p},\nn\\
\partial_z\left(\hat{\psi}_>-\hat{\psi}_<\right)|_{z=z^\p}&=&
\big(1-\xi\big)^{-1}
\frac{1}{k z^\p}
\label{match1}\,.
\end{eqnarray}
By satisfying~(\ref{match1}) and~(\ref{boundaryn}) we can find the
Neumann propagator in $(4+N)$ dimensions. To this end, we follow
the derivations given in the Ref.~\cite{GKR,CNW}. The formula for
the Green function in five dimensions ($N=1$), for both the
arguments of the propagator on the brane, was given, for example,
in~\cite{CNW}, so here we give the expression valid for arbitrary
$N$.
%%%%%%%%%%%%%%%%%%%%% drop this 5d propagator %%%%%%%%%
%\bea &&{\cal G}_5(x,\frac{1}{k};x',\frac{1}{k})\simeq (1-\xi)^{-1}
%\int\frac{d^4 p}{(2\pi)^4}\, e^{ip(x-x')}\nn \\
%&&~~~~~~~~~~~\times \left[-\frac{1-\xi}{1+\xi}\,\frac{2k}{p^2}
%+\frac{(1-\xi)^2}{(1+\xi)^2}\,\frac{1}{k} \ln \left(\frac{ip}{2k}\right)
%\right]\,.\label{propagator}
%\eea
%%%%%%%%%%%%%%%%%%%%%%%%%%%%%%%%%%%%%%%%%%%%%%
The general expression for the Green function in $(4+N)$
dimensions for both the arguments on the brane is \bea &&{\cal
G}_{4+N}(x,\frac{1}{k};x',\frac{1}{k})\simeq (1-\xi)^{-1}
\int\frac{d^{3+N} p}{(2\pi)^{3+N}}\, e^{ip(x-x')}\,
\frac{1}{q}\,\Bigg[\frac{1-\xi}{1+\xi}\,
\frac{N+1}{w}\nn \\
&&~~~~~+\frac{\big(N+(N-2)\,\xi\big)\,
(1-\xi)}{2(N-1)(1+\xi)^2}\,w+ {\cal
O}(w^3)+\cdots+\frac{(1-\xi)^2}{(1+\xi)^2}\, \frac{w^{N}}{c_1}\,
\ln\Big(\frac{w}{2}\Big)\Bigg]\,, \label{propagator2} \eea where
$q/k\equiv w$, $c_1$ is a dimensional constant ($c_1=1, 4,
64,\cdots $, for $N=1, 3, 5,\cdots$), but the last expression
involving a logarithmic term will be absent for even numbers of
extra dimensions ($N=2, 4, 6,\cdots)$. We should mention that the
second expression above will appear only for $N>1$, though we have
written it for arbitrary $N$. Note that only the first two terms
do not vanish for $\xi=1$.

We may split equation~(\ref{propagator2}) into zero mode and KK
mode
\begin{equation}\label{greensum}
\G_{4+N}(x,\frac{1}{k};x', \frac{1}{k})=
\frac{(N+1)k}{1+\xi}\,\int\frac{d^{3+N}p}{(2\pi)^{3+N}}
\frac{e^{ip(x-x^\prime)}}{q^2} +\G_{KK}(x,x')\,,
\end{equation}
where \be \G_{KK}(x,x^\prime) =
-\,(1+\xi)^{-1}\int\frac{d^{3+N}p}{(2\pi)^{3+N}}
e^{ip(x-x^\prime)}\,
\Bigg[\frac{N+(N-2)\,\xi}{2(N-1)}\,\frac{1}{1+\xi}\,\frac{1}{k}
+\frac{1-\xi}{1+\xi}\,\frac{(q/k)^{N}}{q\,c_1}\,
\ln\left(\frac{q}{2k}\right)+\cdots\Bigg]\,. \label{kkmode} \ee
Equation~(\ref{greensum}) reflects a useful relation \be
\label{relate4and5} G_{N+3}=\frac{(N+1)\,k}{1+\xi}\,G_{N+4} \ee
One observes from equation~(\ref{greensum}) that in the low-energy
scale $q/k\,(\equiv w)<< 1$ and for $N>1$ the contribution to the
Newton potential from the continuum KK modes is almost negligible
compared to the contribution of zero mode. One finds $N=1$ as the
simplest example. The limiting behavior of the Newtonian potential
due to a point source of mass $2\pi\,m_*$ on the brane is
therefore~\cite{CNW}
%%%%%%%%%%%%%%%%%%%%%%%%%%%%%%%%%%%%%%%%%%%%%%%%%%%%%%%%%%%%%%%%%%
\bea
V(r)&=&\frac{2\pi m_*}{M_{(5)}^{3}}\int dt\, \G_5\,
\left(r,\frac{1}{k}; 0,\frac{1}{k}\right)\nn\\
&\simeq  & -\frac{G_4\,m_*}{{ r}}\bigg[1+\frac{(1-\xi)}{(1+\xi)}\,
\frac{1}{2\,k^2\,r^2}+\cdots\bigg]\,.\label{onbranepoten} \eea For
$\xi=0$, this expression agrees with the result found
in~\cite{GTT}. A difference of factor $1/2$ to the leading-order
potential correction from that of the original RS work~\cite{RS2}
(which is the case of $\xi=0$) arises from the convention in a
relation between four- and five- dimensional Newton constants. If
one defines~\cite{RS2} \bea \label{RSrelation}
G_4&=& \frac{k}{(1+\xi)}\,G_5 \nn \\
M_{(4)}^2&=&\frac{1+\xi}{k}\,M_{(5)}^3 \label{newtonconst} \eea
instead of relation~(\ref{relate4and5}) the factor of $1/2$ would
be absent. To reconcile the result with an exact graviton
propagator analysis, however, one may have to relate $G_{N+3}$ and
$G_{N+4}$ according to relation~(\ref{relate4and5}). For $\xi=1$,
there is no correction from the continuum KK modes. Moreover, to
avoid anti-gravity effect $(G_4 < 0)$, one requires $\xi > -1$,
and another limit $\xi\leq 1$ arises from the propagator analysis;
this is consistent with the previous observation. Thus, for
$\xi=1$ solution, there are no bulk propagation of graviton in
$AdS_{4+N}$ space. By taking $\xi=0$, from~(\ref{RSrelation}), one
recovers the relation $M_{pl}^2=M_{(5)}^3\,k^{-1}$ and also
recovers the Newtonian potential in a form identical to that
obtained by Randall and Sundrum~\cite{RS2}. For $r >> 1/k$, the
continuum KK modes with $m>> k$, which are normally suppressed on
the brane, would only sub-dominantly contribute to the potential
generated by the $4d$ graviton bound state. For distances $ r<<
1/k$ or at the origin, the continuum modes with $m>>k$ have
unsuppressed wavefunction, thus as anticipated we get the
$(4+N)$-dimensional potential even if $\xi>0$.

A remark is in order. For $\xi=1$ and $N>1$, only the first term
in~(\ref{kkmode}) can survive, but this is negligible as compared
to the zero-mode contribution. One, therefore, always recovers a
localized gravity on the brane which precisely follows the
Newton's law, and there may almost be trivial correction to the
gravitational potential. We will show in next section that $\xi=1$
is indeed a viable solution of the RS-type braneworld models.

\section{Effective solutions with $\xi=1$}

It is not difficult to show that $\xi=1$ explains the RS warped
braneworld compactification. It looks reasonable first to solve
the modified Einstein equations in five dimensions. To this end,
we shall introduce a brane action corresponding to a RS singular
$3$-brane with a positive brane tension $\lambda>0$. Thus the
low-energy effective action in five dimensions can be taken to be
\be \label{thick1} S=\int
d^{5}x\,\sqrt{-g_{5}}\,\left(\frac{R}{\kappa_{5}}
-\Lambda+\alpha\left(R^2-4 R_{pq}R^{pq} +
R_{pqrs}R^{pqrs}\right)\right) +\int d^4x\,\sqrt{|g_4|}\,
(-\lambda)\,. \ee The RS single brane action is recovered for
$\alpha=0$. The modified Einstein field equations following
from~(\ref{thick1}) would simplify to give \bea
\left({A^\p}^2+A^{\p\p}\right)\left(1-\xi \right)&=&
\frac{\lambda}{6}\,\kappa_5\, \delta(z)
\label{constphi1}\\
{A^\p}^2 -\frac{\xi}{2}\,{A^\p}^2
&=&-\frac{\kappa_5}{12}\,\Lambda\,e^{-2A(z)} \label{constphi2}
\eea where $\xi\equiv 2\epsilon\,e^{2A(z)}{A^\p}^2$, and
$\epsilon\equiv 2\alpha\kappa_5$. One might have already noted
that appealing to $\xi=0$ in the bulk essentially implies
$A^{\p\p}+{A^\p}^2=0$. In fact, $\xi=0$ kills another possible
solution, i.e. $\xi=1$, which indeed gives a physical solution. As
is known, any braneworld compactification with $\xi=0$ suffers
from the so-called {\it c-theorem}~\cite{Freedman99a,Malda00a}
based on Einstein theory. We should mention that the {\it
c-theorem} may not be available to $\xi=1$ solution. We will
explain this below only briefly, and detail discussions on this
issue will appear elsewhere.

One clearly sees from~(\ref{constphi1}) that
$\xi\equiv 2\epsilon {A^\p}^2
e^{2A(z)}=1$ is a bulk solution of the field equations.
From this we find \bea \label{solution}
e^{A(z)}&=& \int_0^{z}\frac{1}{\sqrt{2\epsilon}}\,dz \nn\\
e^{-A(z)}&=& \frac{\sqrt{2\epsilon}}{|z|+\sqrt{2\epsilon}}\equiv
\frac{l}{|z|+l}\,, \eea where we have normalized the solution so
that $A(0)=0$. Since we have a non-trivial brane tension
$\lambda>0$ at the brane $z=0$, equation~(\ref{constphi1}) implies
that we must satisfy the following relation on the brane $z=0$:
\be \frac{12\delta(z)}{l+|z|}\,\left(1-\frac{2\epsilon}{L^2}\,
sgn(z)^2\right)=\lambda\,\kappa_5\,\delta(z)\,. \ee This, after
regularizing the $\delta$-function, uniquely determines the
$3$-brane tension, \be \label{tension1} \lambda=\frac{1}{2\pi
G_5\,l}\,. \ee Next, equation~(\ref{constphi2}) can be used to fix
the bulk cosmological constant \be \Lambda=-\frac{3}{8\pi
G_5\,l^2}\,. \ee The readers can easily check that $\xi=1$
consistently gives a physical solution even if the number of extra
dimensions $N>1$. It is interesting that $e^{-A(z)}$ will converge
as $z\to \infty$, this is what we need for the solution to be
physical. If one interprets the singularity at $A^{\p\p}+{A^\p}^2$
as the infrared cut-off (or the RS singular brane), then we may
interpret the $z\to \infty $, $\xi=1$ as the ultraviolet cutoff or
anti-de Sitter boundary where gravity is strong. Changing to
$y$-coordinate $dy=e^{-A(z)}\,dz$, the weak energy condition reads
$A^{\p\p}(y)\geq 0$, and the warp factor reads $A(y)=\pm |y|/l$.
Since $A^\p(y)$ takes a positive (negative) value for $y\to
+\infty~(-\infty)$, the {\it no-go theorem}~\cite{Freedman99a} may
not apply to the $\alpha>0$ case.

\section{Completely localised gravity with $\xi=1$}

As we already emphasized by considering $\xi= 1$, we can obtain
equivalent $N$ copies of the RS-type non-relativistic
Schr\"odinger equations, but in one lower spacetime dimensions. We
shall express the metric fluctuations again in a canonical form by
changing the variables in equation~(\ref{lineareq1}) as
$h_{\mu\nu}=e^{(N+1) A(z_i)/2}\,\hat{\psi}(z_i) e^{ip\cdot x}\,
\epsilon_{\mu\nu}$. Then the non-trivial part of linearized
equations takes the following form
\begin{eqnarray}
&&8 N\alpha\,\kappa\,k\,
\sum_{i=1}^{N}\delta(z_i)\bigg[\bigg(-m^2-\sum_{i\neq j}^{N}
\partial_{z_j}^2
-(N+1)ke^{-A(z)}\sum_{i\neq j}^{n}\delta(z_j)\nn \\
&&~~+ \frac{(N-1)(N+1)(N+3)\,k^2}{4(k\sum_{i\neq j}|z_j|+1)^2}\bigg)\,
\hat{\psi}(z)+(N+1)\,sgn(z_i)\,\Big(\partial_{z_i}
+\frac{(N+1)}{2}\,A'(z)\Big)\,
\hat{\psi}(z)\bigg]=0\,.\nn \\
&&
\label{lineareq3}
\end{eqnarray}
Equation~(\ref{lineareq3}) clearly implies that the so called
'solitonic' solutions defined for $\xi= 1$ have the structure
which is formally the same as the ``normal'' or RS-type brane
solutions, other than the last term. Some of the clear differences
here are that (i) there is no bulk propagation of the graviton in
$(4+N)$-dimension, (ii) the {\it solitonic} brane is
$\delta$-function like, which is also seen from the overall
coefficient in~(\ref{lineareq3}). The last term in
equation~(\ref{lineareq3}) implies an appropriate jump
condition(s) to be satisfied across the branes at $z_i=0$, but
does not contribute for $z>0$. Thus, for $\xi= 1$, the
$(N-1)$-dimensional Schr\"odinger equation can also be written in
the form \be
\frac{4\xi\,\delta(z_i)}{N(N+1)k}\,\left[-\partial_{z_j}^2+
\frac{(N+1)^2}{4}
{A^\prime}^2-\frac{N+1}{2}A^{\prime\prime}\right]\hat{\psi} =m^2
\hat{\psi}\,, \label{solitonbulk1} \ee where $A=\log
\big(k\sum_p|z_p|+1\big)$ and $i, j=1, 2\cdots, N, ~ i\neq j $.
Thus, the bulk equations along $z_j$-directions are \be
\frac{4\xi\,\delta(z_i)}{N(N+1)k}\,\left[-\partial_{z_j}^2+
\frac{(N-1)(N+1)(N+3)}{4\,(|z_j|+1/k)^2}\right]\hat{\psi} =m^2
\hat{\psi}\,. \label{solitonbulk2} \ee

%%%%%%%%%%%%%%%%%%%%%%%%%%%%%%%%%%%%%%%%%%%%%%%%%%%%%%%%%%
The simplest choice is $N=2$. Then~(\ref{solitonbulk2})
gives two five dimensional bulk equations of graviton, for each
$i,j=1,2$ but $i\neq j$. For the continuum
of KK modes propagating along the solitonic $4$-branes located at the
$z_1$ and $z_2$ axes, the continuum mode solutions are given by a
linear combination
\be
\hat{\psi}_m^{(j)}(z_j) \sim N_m^{(j)}\,\sqrt{(|z_j|+1/k)}\,
\Big[J_2\big(m(|z_j|+1/k)\big)
+B_m^{(j)}\,Y_2\big(m(|z_j|+1/k)\big)\Big]\,,
\ee
where $j=1,2$. The coefficients $B_m^{(j)}$ and $N_m^{(j)}$ are
determined from the boundary conditions and $\delta$-function
normalization condition. These are readily evaluated to be
\be
B_m^{(j)}=-\frac{Y_1 (m/k)}{J_1 (m/k)}\,,
 ~~ N_m^{(j)}=\sqrt{\frac{m}{4\big(1+{B_m^{(j)~2}}\,\big)}}\,.
\ee We consider the contribution of KK graviton exchange to the
non-relativistic gravitational potential between two point masses
$m_1,\,m_2$ placed on the intersection of two orthogonal solitonic
$4$-branes with a separation of $r$. A total contribution of the
Newtonian potential is therefore \bea
-\frac{V(r)}{m_1\,m_2}&=&\left[\frac{G_4}{r}+\frac{3k}{2}\,M_*^{-4}
\int_{m_0}^{\infty} dm\,|\hat{\psi}_m^{(j)}(0)|^2\,
\frac{e^{-mr}}{r}\right]\nn \\
V(r)&=&-G_4\,\frac{m_1\,m_2}{r}\left[1+\frac{1}{k^2\,r^2}\right]\,.
\eea where $G_4=(3/4)\times k^2M_*^{-4}$ is used. In arriving at
the second line it was important to use $|\hat{\psi}_m(0)|^2\sim
m/(4k)$, which one can easily find from plane wave normalization
conditions of the Bessel functions as in~\cite{Csaki00a}.

For $N=3$, equation~(\ref{solitonbulk2}) would rise to give the
following three six-dimensional bulk equations of the graviton,
each bulk mode propagating along the solitonic $5$-branes located
at $u(=z_p-z_q)$ and $v(=z_p+z_q)$ axes. Hence
 \be
\frac{2\,\delta(z_i)}{3k}\,\left[-\partial_{u}^2-\partial_{v}^2+
\frac{6}{(|v|+1/k)^2}\right] \hat{\psi}(u,v) =\half\, m^2
\hat{\psi}(u,v)\,, \ee where $i=1, 2, 3$, and $i\neq p \neq q$.
In order to satisfy the appropriate boundary condition(s) implied by
the $\delta$-function potential at $z_i=0$, we must choose
the linear combinations \bea
\varphi_{m}(u)&\sim& N_1(m)\,\sqrt{u}\,
\big[\sin(m_u u)+A_m\cos(m_u u)\big]\\
\varphi_{m}(v) &\sim& N_2(m)\,\sqrt{(|v|+1/k)}\,
\Big[J_{5/2}\big(m_v(|v|+1/k)\big)
+B_m\,Y_{5/2}\big(m_v(|v|+1/k)\big)\Big]\,.\nn \\
&{}&
\end{eqnarray}
These solutions must satisfy the Neumann type boundary conditions
at the brane junction $u=0$, $v=0$: \be
\left(u\,\frac{\partial_{v}- \partial_{u}}{2}+\frac{5}{2} \right)
\hat{\psi}=0\,,\quad \left(v\,\frac{\partial_{v}+
\partial_{u}}{2}+\frac{5}{2} \right) \hat{\psi}=0\,. \ee For
$m_u^2,\,m_v^2>0$, satisfying the boundary conditions, the
coefficients $A_m$ and $B_m$ are determined to be \be
A_{m_u}=\cot(m_u u)\,,
~~B_{m_v}=-\frac{Y_{3/2}(m_v/k)}{J_{3/2}(m_v/k)}\,. \ee The
continuum wavefunction therefore reads \be
\hat\psi_{m\,{(j)}}=\varphi_m(u_j)\times \varphi_m(v_j)\,. \ee By
considering the contribution of KK graviton exchange between two
unit point masses $m_1,\,m_2$, placed on the intersection of three
intersecting $5$-branes with a separation of $r$, one may find the
total contribution of the Newtonian potential. This is given by
\bea -\frac{V(r)}{m_1\,m_2}&\simeq &\frac{G_4}{r}+\frac{3k^2}{2}\,
M_*^{-5}\int_{m_0}^{\infty}
dm_u\,dm_v\frac{e^{-\widehat{m}r}}{r}\,
|\hat{\psi}_m^{(j)}(0)|^2\nn \\
V(r)&\simeq
&-G_4\,\frac{m_1\,m_2}{r}\left[1+\frac{c_1}{(k\,r)^3}\right]\,.
\eea where $\sqrt{m_u^2+m_v^2}\equiv \widehat{m}$,
$G_4=(3/2)\times k^3M_*^{-5}$ and $c_1$ is some constant of order
$1$. Again in arriving at this result, we have used the value
$|\hat{\psi}_m(0)|^2\sim m_v^2/(6k^2)$ determined from the plane
wave normalization conditions of the Bessel functions, and
specialized to the case where $m_u$ would extend down to $m_u=0$,
and $\psi_{m_u}(0)=1$.
%%%%%%%%%%%%%%%%%%%%%%%%%%%%%%%%%%%%%%%%%%%%%%%%%%%%%%%%%%%%%%%%%%%%%%%%%

\section{Comments on scalar and vector modes}

So far we have analysed the tensor mode $h_{\mu\nu}$ of the metric
perturbations, leading to graviton propagators and corrections to
Newton potential. This analysis was important because a
high-dimensional graviton may represent a continuum of
four-dimensional states and the gravitational potential on the
brane is mediated by an effective four-dimensional graviton. It
may be relevant to know whether the scalar/vector modes of the
metric perturbations have any significant roles in the higher
dimensional brane background. One of the motivation to look after
these modes is that in the conventional Kaluza-Klein theory of
compact extra spaces, because of the isometries of a factorizable
geometry, scalar (vector) modes correspond to the physical
gravi-scalar (gravi-photon) modes~\cite{Salam81}. But this does
not seem to be the case for a non-factorizable geometry.

The scalar and vector modes of the metric fluctuations, without
any gauge choice, satisfy \bea
\left(1-\xi\right)\Big[\partial^\mu\partial^\nu\left(h_{\mu\nu}-\eta_{\mu\nu}
h\right)+(N+2)\,\partial_z A\left(\partial_z h-2\partial^\lambda
h_{\lambda z}+(N+3)\,\partial_z A\,h_{zz}\right)\Big]&=&0\,.\label{zzcompo} \\
\left(1-\xi\right)\left[\partial_z\left(\partial^\lambda
h_{\mu\lambda}-\partial_\mu h\right)- (N+2)
\partial_z A\,\partial_\mu h_{zz}+\partial^\lambda
\partial_{[\mu} h_{\lambda]z}\right]&=&0\,.\label{muzcompo} \eea
Here $h\equiv h_\lambda^\lambda$,
$\partial_zA=A^\p\,\partial_z|z|=k\left(2\Theta(z)-1\right)
=k\,\sum_{i=1}^N\,sgn(z_i)$, and
$\partial_z=\partial_{z_1},\cdots, \partial_{z_N}$. (These
equations were reported in~\cite{IPN01a,CNW} but for $N=1$ and in
the axial gauge, i.e., $h_{zz}=0=h_{\mu z}$).
Equations~(\ref{zzcompo}, \ref{muzcompo}) are trivially satisfied
for $\xi=1$ solution. There are no terms having second derivatives
of the warp factor $A(z)$, and hence there are no delta-function
sources for scalar/vector modes. This is not the case for the
tensor mode $h_{\mu\nu}$ (see, for example,
equation~(\ref{linearxi})), even after the setting $\xi=1$. Thus
the scalar and vector modes of the metric fluctuations alone do
not appear as dynamical fields of the
theory~\cite{RS2,CZK,IPN01a}, and hence may not be localized on
the brane. This result is consistent with the observations made
in~\cite{CZK,Giovannini01a}.

Since a massless graviton on the brane is assumed to be transverse
and tracefree, it is reasonable to impose the gauge
$h_\mu^\mu=0=\nabla^\nu h_{\mu\nu}$ to simplify the
expressions~(\ref{zzcompo}), (\ref{muzcompo}). For $N=1$, since
$Z_2$ symmetry on the brane already fixes $h_{\lambda z}=0$, one
finds $h_{zz}=0$, thus the graviscalar and graviphoton components
are pure gauge degrees of freedom. If the brane gravity is coupled
to brane matter, one may still gauge them away using the
brane-bending formalism discussed
in~\cite{GTT,GKR,CNW}~\footnote{See, for example,~\cite{Ruba00a}
for related comments about the cancellation of the scalar mode by
the 'brane-bending' mode in the framework of a five dimensional
$3$-brane model. These comments may not be directly applicable to
the $3$-brane defined as the common four dimensional intersection
of higher dimensional branes.}.

One can, of course, couple the scalar modes of the metric
fluctuations to the scalar fluctuation of a non-trivial bulk
scalar field. But the effective theory will be of scalar-tensor
nature rather than conventional tensor gravity. Unlike the tensor
modes, the scalar/vector modes of the metric perturbations are not
automatically invariant in a gauge invariant manner. Indeed, for
$N>1$, one may not set all the scalar/vector modes to zero or
gauge them away, and in principle one can analyse them using some
gauge-invariant formalism. The analysis is relatively simpler in a
RS-type warped brane background with one extra dimension (i.e.
$N=1$)~\cite{Giovannini01a,Bozza01a}, but the diagonalisation of
linearized fluctuations of all metric fields for $N>1$ is too
hard, if not impossible. Some progress can be made by taking the
simplest choice, $N=2$, and using the gauge invariant formalism
developed in~\cite{deBruck00a}. We take the viewpoint that
consideration of scalar and vector components would not change the
results, at least the momentum structures of the graviton
propagators, of this paper.

\section{Conclusion}
%%%%%%%%%%%%%%%%%%%%%%%%%%%%%%%%%%%%%%%%%%%%%%%%%%%%%%%%%%%%%%%%%%%%%%

We have studied gravity localization to the RS branes for a more
general class of $N$ intersecting $(2+N)$ brane configurations by
including higher order curvature terms in a Gauss-Bonnet
combination. For $N= 2$, a GB term can support a feasible
singularity at the brane junction, thereby allowing a non-trivial
brane tension at the four-dimensional brane intersection. But this
would not have been possible if the braneworld action had a
contribution only from the Einstein term and the cosmological
constant term~\cite{Hamed}. The effective four-dimensional Planck
scale $M_{Pl}$ has been derived, and also the explicit expressions
for the graviton propagator in $(4+N)$ spacetime dimensions are
obtained. We have explained in some detail interesting features of
localized gravity with a Gauss-Bonnet term for 'normal' or RS-type
branes supported by $\delta$-function-like sources, and also
so-called {\it solitonic} branes defined for the class of solution
$\xi=1$.

The present analysis might reveal some important features of the
localized RS type braneworld solutions or {\it solitonic}
braneworld gravity created by non-dynamical sources (i.e. in the
absence of fields other than gravity). We have shown that the RS
single brane model with a Gauss-Bonnet term in the bulk correctly
gives a massless graviton on the brane as for the RS model. We
provided a complementary description of the Newton potential
corrections to the long-range interaction due to continuum modes
living in the AdS space. Basically, our work extends the previous
work in the literature~\cite{Hamed,GKR,CNW} to the general (4+N)
dimensional anti-de Sitter spacetimes, but the results reported
here also include the contribution of the Gauss-Bonnet interaction
term. The latter is the only viable combination of higher
curvature corrections that one can introduce in the RS braneworld
scenario of warped extra dimensions.

For the class of solution $\xi=1$, the scalar and vector fields do
not appear as the propagating degrees of freedom in the transverse
$z$ direction, though the zero mode of the tensor fluctuation is
localised on the brane. This reveals that the physical brane could
be a solitonic $3$-brane living in six-dimensional spacetimes,
rather than in five-dimensional AdS space. The number of
non-compact extra dimensions, $N=2$, is itself more interesting in
the scenario of~\cite{Hamed,Hamed1}.

In nutshell, the physical relevance of our analysis is two-fold --
it is a meaningful extension of the RS model with an extra
non-compact dimension to an arbitrary $N$, and -- a natural
generalization of the ADDK~\cite{Hamed} scenario of infinite large
new dimensions including the contribution of a Gauss-Bonnet term
into the effective braneworld (gravitational) action. The present
analysis also provides hints for a possible creation of {\it
solitonic} $3$-brane in a six dimensional warped bulk background,
which would not be possible without the higher curvature terms. It
is plausible that the solitonic braneworld solution defined with
$\xi=1$ in $N\geq 1$ has a structure which is formally the same as
for 'normal' or RS-type solutions.

%\vspace{0.5cm}
\section*{Acknowledgments}

I would like to thank M. Blau, Gregory Gabadadze, N. Kaloper and
Carlos Nunez for fruitful discussions and correspondences. This
work was supported in part by Seoam Foundation, Korea, and by the
BK21 Project of the Ministry of Education. I would like to
acknowledge a kind hospitality of Abdus Salam ICTP, where the
manuscript was revised.

%\begin{references}

\end{document}